# A new variable for SRS plan quality evaluation based on normal tissue sparing: The Effect of Prescription Isodose Levels

Q. Zhang, D. Zheng, Y. Lei, B. Morgan, J. Driewer, M. Zhang, S. Li, S. Zhou, W. Zhen, R. Thompson, A. Wahl, C. Lin and C. Enke

*Department of Radiation Oncology*

*University of Nebraska Medical center, Omaha, Nebraska 68198*

ABSTRACT

**Objectives:** A new dosimetric variable, dose dropping speed (DDS), was proposed and used to evaluate normal tissue sparing among stereotactic radiosurgery (SRS) plans with different prescription isodose lines.

**Methods:** Forty plans were generated for 8 intracranial SRS cases, prescribing to isodose levels (IDLs) ranging from 50% to 90% in 10% increments. Whilst maintaining similar coverage and conformity, plans at different IDLs were evaluated in terms of normal tissue sparing using the proposed DDS. The DDS was defined as the greater decay coefficient in a double exponential decay fit of the dose drop-off outside the PTV, which models the steep portion of the drop-off. Provided that the prescription dose covers the whole PTV, a greater DDS indicates better normal tissue sparing.

**Results:** Among all plans, the DDS was found the lowest for the prescription at 90% IDL and the highest for the prescription at 60% or 70%. Beam profile slope change in penumbra and its field size dependence were explored and given as the physical basis of the findings.

**Conclusions:** A variable was proposed for SRS plan quality evaluation. Using this measure, prescriptions at 60% and 70% IDLs were found to provide best normal tissue sparing.

**Advances in knowledge:** A new variable was proposed based on which normal tissue sparing was quantitatively evaluated, comparing different prescription IDLs in SRS.

**Short title: SRS plan normal tissue quality evaluation**

## I. INTRODUCTION

Stereotactic radiosurgery (SRS) has gained increasing popularity as a treatment modality for patients with brain metastases as well as other malignant and benign brain lesions.[1] SRS has traditionally been performed by using an invasive fixed head frame that establishes the stereotactic coordinates of the target.[2] More recently, frameless stereotactic systems have been developed and implemented with the help of an image-guided system.[3-11]

The reports of radiation therapy oncology (RTOG reports[12-15]) have made specific prescription dose recommendations for brain SRS treatments based on different target volumes. But the prescription isodose level (IDL) can vary from 50% to 90% among different clinical practices. Therefore, it is interesting to find out which prescription IDL would be most suitable for brain SRS. In fact, a recent study has been conducted in this aspect[16]. In Ohtakara et al's study [16] 10 SRS cases have been retrospectively planned and studied comparing different IDLs (90%, 80%, and 70%), and the authors have found the best prescription IDL at 70% for those 10 cases, based on $V_{50\%}$. The technique used in their study was non-coplanar dynamic conformal arcs, standard in brain SRS treatments. However, in the study, no physical reason was explored to explain the findings, and the studied prescription IDLs ranged from 70% to 90%, rendering it inadequate to determine if 70% was truly the extrema. Whilst we explored a similar topic in this paper, we used a broader search range of 50%-90%, the range of clinically used prescription IDLs, to comprehensively study the normal tissue dose effect of prescription IDLs. Furthermore, we proposed in this work a new and useful variable, dose dropping speed (DDS) which reflects the radial dose drop-off from the PTV surface, defined as the greater decay coefficient in a double exponential decay fit of the dose drop-off outside the PTV, to quantitatively evaluate the normal tissue sparing. The double exponential decay fit takes a global look at the dose drop-off outside the PTV, with the greater decay coefficient characterising the steep portion of the drop-off or the higher dose gradient region, and the lesser decay coefficient characterising the shallow portion or the lower dose gradient region. Our work chose to define the greater decay coefficient, due to its greater clinical relevance, as a quantitative measure called DDS, and used it in our investigation of normal tissue dose effect of prescription IDLs. In addition to discovering the effects of the prescription IDLs on the plan quality, our work also explored the physical aspects in the attempt to explain the observations. The normal tissue sparing trend of different prescription IDLs was found to result from the different gradients of the penumbra on the LINAC beam profile of the corresponding effective field size. Effectively speaking, a different part of the beam penumbra of a different beam field size is used to surround the PTV on the plan of a given prescription IDL. Moreover, our work also uncovered a target size dependence of the observed normal tissue dose trend and explored its physical basis. Finally, the planning technique had also some difference from Ohtakara et al's work. In their study, block margin was only uniformly adjusted for plans with different prescription IDLs, whilst in our study, we manually optimised individual MLC leaf positions for each plan to generate realistic plans with quality satisfactory to our clinical standards. As a result, in our work the plans with different IDLs for the same patient all had high and matched coverage and conformity, therefore rendering our results and conclusions more generalizable to clinical SRS practice.

It is difficult to simply define what a good plan entails. But in general, a good plan shall have both good local control and at the same time good normal tissue sparing to the extent possible, and this is especially true for SRS cases in which an ablative dose is given in a single fraction. In other words, hot spots in the PTV may be tolerated but the normal tissue shall be as cold as possible. As will be shown in this paper, the prescription IDLs were revealed to affect the normal tissue sparing in an SRS plan. Normal tissue sparing is usually quantitatively or qualitatively evaluated in a variety of ways. One way is to assess specific dose-volume endpoints for different organs at risk (OARs), such as the maximum dose to the brainstem, or the volume of normal brain tissue getting over 12 Gy. These endpoints are useful because they are often linked with specific found toxicities and therefore represent what the clinicians are most interested in to constrain the dose and minimise the toxicities. On the other hand, each of these measures only represents a local and partial view of the entire plan. As a result, those variables cannot be used to represent the general dose change trends inside the normal tissue. Another popular way to evaluate normal tissue dose is to use $V_{50\%}$ as has been done in the study of Ohtakara et al [16]. However, $V_{50\%}$ has the same shortcoming as the dose volume histogram that it does not reflect the "coordinate information". In other words, for the same $V_{50\%}$ the dose can be distributed in many different positions. Furthermore, $V_{50\%}$ only studies the dose effect at a localised dose level, i.e. 50%. Finally, the selection of $V_{50\%}$ not $V_{55\%}$ or $V_{40\%}$ is arbitrary. Thus there is a natural question whether we could find a generic function which can be used to describe the dose distribution outside the PTV. In this work, we have proposed a double exponential function for this purpose. In this work we proposed a new parameter named DDS, extracted from a global fitting of the dose drop-off outside the PTV, to quantitatively and comprehensively evaluate normal tissue sparing, and study the effects of different prescription IDLs in brain SRS.

Our dose drop-off calculation involves some position information of the dose distribution which is ignored in the above-enumerated measures, and thus provides us information which can be used to evaluate the general dose trends outside the PTV. The DDS as we defined using the greater decay coefficient in a double exponential decay fit of the dose drop-off outside the PTV is of course an interesting measure which has not been explored in the existing literatures.

The goals of SRS are the ablation of target tissue and the sparing of critical normal tissue. Largely due to the little normal tissue usually contained in an SRS PTV, dose inhomogeneity inside the PTV is considered acceptable. Conventionally, the "sphere-packing" type of SRS, such as Gamma Knife, prescribes to a fairly low IDL, usually around 50%, whilst such low prescription IDL would not be used for radiotherapy with conventional fractionations in which target uniformity is critical. High prescription IDLs such as 90% are achievable in LINAC-based SRS, although a broad range of IDLs from 50% to 90% have been used in clinical practice due to the relative freedom from the target uniformity restriction and the importance of normal tissue sparing outside the PTV. Whilst our work was conducted based on our proposed DDS to study the normal tissue dose effect and identify the prescription IDL that would provide the optimal normal tissue sparing outside the PTV, there are a few other items need to be taken into consideration. Most important of all, the tolerance of target dose heterogeneity is still clinically important even for SRS. Although usually within the target itself, the hottest dose point in a low prescription IDL plan can have much higher dose than that in a high prescription IDL plan. For example, for a plan with 18 Gy prescribed to 90% IDL, the hottest dose is around 20 Gy whilst another plan with the same dose prescribed to 50% IDL will have the hottest dose at around 36 Gy. Because single fraction SRS treatment was found to cause necrosis sometimes[17-20], very high dose points shall still be avoided and therefore plans with very low prescription IDLs may not be clinically practical or appropriate. In addition, because plans with lower prescription IDLs deliver higher maximum doses and require larger numbers of monitor units, they may in turn result in larger integral dose and may cause logistic problems such as longer irradiation time.

Last but not the least, a plan with a lower prescription IDL has a smaller effective field size than a plan with a higher IDL. Because the typical smallest multi-leaf collimator (MLC) field size commissioned for LINAC-based SRS is about 1 cm, the smaller effective field size in low IDL plans for a very small target may lead to bigger dosimetric errors than a larger field size in high IDL plans would.

Here we report our work in which we used the DDS calculated from a double exponential decay function fit of dose drop-off outside the PTV to investigate normal tissue sparing on brain SRS plans with 50%-90% prescription IDLs. A total of 40 plans were retrospectively generated for 8 brain SRS patients who had been clinically treated with 90% IDL plans. The physical basis of the observed phenomena will also be discussed.

## II. METHODS

Under a study approved by the Institutional Review Board, eight previously treated brain SRS patients were randomly selected, including one acoustic neuroma, one meningioma, two pituitary lesion, and four metastatic tumors. Table 1 lists the disease sites, locations, and the PTV volumes for all the patients. These patients were clinically treated prescribing to 90% IDL following the practice guideline at our institution. In our work, a single physicist first re-planned the 8 cases at 90% IDL with non-coplanar 6 MV dynamic conformal arcs using iPlan (BrainLab AG, Feldkirchen, Germany) for a Novalis machine (BrainLab AG, Feldkirchen, Germany). Subsequently, whilst fixing the arc number and orientation for each patient, the physicist manually optimised the individual microMLC leaf positions to create plans with the other prescription IDLs (80%, 70%, 60%, and 50%). The plans were created such that the PTV coverage and conformity were similar for the same patient on the different plans. The typical number of arcs was around five.

To quantitatively evaluate the dosimetric effect on the normal tissue for individual plans, we proposed to use a new metric, the DDS, defined as the following. Firstly, one-mm-thick concentric rind structures were generated layer by layer from immediately outside the PTV to when the rind structure reaches the head surface (see an example patient image in Fig. 1). Secondly, the average dose inside each rind structure was calculated. Thirdly, an analytical double exponential decay function was fitted to describe the average dose as a function of the distance from the PTV surface, as in Eq. (1).

$$D(r) = a_1 \exp(-b_1 r) + a_2 \exp(-b_2 r) \qquad (1)$$

where D(r) [Gy] denotes the average dose in a rind structure with the distance of r [mm] from the PTV surface. The two terms are symmetrical in the Eq. (1). The steeper decay is always denoted as the first exponential term $a_1 \exp(-b_1 r)$, and the shallower decay as the second exponential term $a_2 \exp(-b_2 r)$. Lastly, we defined $b_1$, the greater decay coefficient from the steeper decay of the fit, as our proposed variable DDS. When $r \to 0$, which corresponds to the surface immediately outside the PTV, the second term of Eq. (1) approaches a constant value. Therefore, the first exponential dominates the dose drop-off in the areas outside but close to the PTV, i.e., the medium-to-high dose region outside the PTV. On the other hand, at a point farther away from the PTV surface, the contribution of the first term becomes less prominent. For example, at a point with the distance of $\frac{\ln 2}{b_1}$ mm from the PTV surface, the contribution of the first term is $\frac{a_1}{2}$.

At a point with the distance of $n \cdot \frac{\ln 2}{b_1}$ mm, the contribution of the first term becomes $\frac{a_1}{2^n}$. It is obvious that at $r_0 = \frac{\ln(a_1) - \ln(a_2)}{b_1 - b_2}$, the two terms in Eq. (1) are equal. At a point closer than $r_0$, the first term dominates the contribution, and on the other hand, at a point farther away than $r_0$, the second term dominates. We defined $b_1$ to be our proposed variable DDS, because the dose drop-off in the normal tissue immediately outside and close to the PTV is always of the most clinical importance in brain SRS.

Suppose a plan is made perfectly such that the prescription IDL completely covers and perfectly conforms to the surface of the PTV, a larger DDS is then preferable because it means faster dose drop-off immediately outside and close to the PTV. The OARs close to the PTV will receive lower dose. On the other hand, a smaller DDS indicates slower dose drop-off leading to a higher dose to the OARs close to the PTV.

For the 40 plans we generated for the 8 SRS patients (5 plans with prescription IDLs set at 50%-90% for each patient), we calculated the dose distribution outside the PTV and applied the above described fitting to it for each plan to investigate the normal tissue dose effect of different prescription IDLs. To explore the physical basis behind the studied effect and its trend, the MLC-collimated square field beam profiles measured at the commissioning and modeled by the treatment planning system were inspected in connection with the clinical plans.

To explore the target size dependence we conducted a simulated phantom study in which a hypothetical spherical target was used for brain SRS. In the study we fixed the prescription IDL at 80% and changed the spherical target diameter from 4.90, 4.05, 3.20, 2.30, to 1.2 cm. A plan was generated for each target size, from which the DDS for each target size was calculated.

## III. RESULTS

In Fig.2, one example of the dose as a function of the distance from the PTV surface is shown (Patient 4 in Table 1). In this figure, the prescription IDL is 80%. Using Equation (1) to fit it, the fitted coefficients were

$a_1 = 18.43 \pm 0.57 (Gy),$    $b_1 = 0.2477 \pm 0.013 \ (1/mm),$
$a_2 = 4.218 \pm 0.617 \ (Gy),$    $b_2 = 0.04234 \pm 0.0054 (1/mm)$

For this plan, the halfway dose decay distance for the DDS was $\frac{\ln 2}{b_1} = 2.8mm$. The halfway dose decay distance for the second exponential term was $\frac{\ln 2}{b_2} = 1.6cm$. At $r_0 = 7.2mm$, the contributions from the two terms were equal.

Comparing the 5 different prescription IDL plans for each patient, similar trends were found for all 8 patients. The isodose distributions outside the PTV for one example patient are given in Fig.3 for cases of prescription IDLs from 90% to 50% (Patient 2 in Table 1). From the figure the following two observations were made: (1) It appeared that for the plans with all other prescription IDLs the high dose distribution more tightly hugs the PTV than the 90% plan. (2) For the lower prescription IDL cases (i.e. 50%) the lower dose was more spread out. The average dose in each rind structure as a function of the distance from the PTV surface is plotted for each plan in Fig. 4 for this example patient (Patient 2). It was clear that for the 90% prescription IDL plan, the dose drop-off in the medium-to-high dose region was the slowest. On the other hand, the tissue dose at the largest studied distance from the PTV was the highest for the 50% prescription IDL plan. These findings were in accordance with the observations from Fig. 3.

To quantify the observations from Fig.4, Eq. (1) was used to fit the dose distribution and the results are provided in Table 2. The conformity index, defined as the ratio of the prescription dose covered volume to the PTV volume, is also given in Table 2 for all plans. It was clear that similar conformity was achieved across different prescription plans for the patient. Because the first rind in the data extraction was so close to the PTV (1mm from the surface of the PTV) that its average dose was often more affected by the minute differences in plan conformity, we chose to exclude this point from the fitting, which led to the fitted coefficients shown in row 2 to row 5. The total MUs for all plans are also provided in Table 2. It is interesting to note that the DDS was found optimal (the highest) in the 70% plan. For other patients not shown, the DDS was found optimal either in the 70% or the 60% plan. The exact optimal percentage IDL may depend on the PTV volume, location, and other factors, though all 8 patients showed a plateaued DDS peak at around 60-70%. The other observation is that for a lower IDL, the MU number is also increased and this is understandable. For a lower IDL, the maximum dose inside the PTV increases, thus the corresponding MU also increases. The numbers of MU of the plans with different IDLs are almost inverse proportional to the IDLs. For example, the MU ratio for the plans at 50% and 90% IDL is 1.838, which is almost the same as the inverse ratio of the two IDLs which is 1.8. We need to point out that in addition to the higher dose heterogeneity inside the PTV, the higher MU at the lower IDL may also lead to slightly broader low dose region, although our DDS study concentrates on the more clinically-relevant region of medium-to-high dose.

The DDS values of different prescription IDL plans of all eight patients are presented in Fig.5. From the plot, it is clear that the DDS was always the lowest in the 90% IDL plan and it increased with lower IDL plans until it plateaued at about 60-70% prescription IDLs. The ratio of the highest DDS to that in the 90% case was between 1.18 and 2.10. The mean ratio was 1.65±0.33 (one standard deviation). An Anova analysis indicated that the DDS difference was significant between the optimal plans (60% or 70%) and the 90% plans with p<0.01.

In exploring the beam profiles, the observed DDS trend in the clinical cases was found to correlate with the beam profiles in the following ways. The beam aperture size or MLC block margin is different when we prescribe at the 90% IDL versus at 70% IDL. In general, the lower the prescription IDL, the smaller are the beam aperture sizes for the same case. To understand the observations of the effects of prescription IDL on the DDS, our investigation on the beam profiles for different field sizes are presented here.

Fig.6 plots the beam profiles for different MLC-defined square field sizes with varying sizes from 6 mm×6 mm to 52 mm×52 mm. It is apparent that the beam profile became narrower and dropped faster around the penumbra region when the field size was smaller. To more clearly appreciate this, the absolute value of the gradient, calculated using the first-order derivative of the profiles, as a function of the distance from the central axis is plotted in Fig.7. It is interesting to note that the gradient was larger as the field size got smaller.

A target size dependence was found for the DDS from the simulated phantom study. The DDS of the 80% IDL plan was compared for a simulated spherical target of varying diameters. In Fig.8, the DDS as a function of the target diameter is plotted. It is clear that when the target became smaller, the DDS became larger.

A similar trend with varying target sizes was also seen for all but two patient cases. Fig.9 plots the optimal (maximum) DDS from all 5 plans with different prescription IDLs for each patient as a function of the corresponding PTV volume. Except Patient 7 and Patient 8 who had PTV volumes smaller than 1 cc, for the larger PTVs the DDS decreased as the PTV volume increased. This trend as seen in the 6 patients with larger PTVs and in the simulated target study could be due to the similar volume (field size) dependence of the calculated slope of the beam profiles (using the first-order derivatives) as a function of percentage dose of the central axis (CAX) values as plotted in Fig.10. For the two small PTVs that didn't follow the trend, one was very close to the surface of the head (Patient 8) and the other was an acoustic neuroma sitting inside bony structures (Patient 7). For the acoustic neuroma case, the target location was quite different from the others. Therefore, the DDS might have also depended on the target location and its distinct surrounding structures. For Patient 8 for whom the target was close to the surface of the head, in addition to the beam profile effects described above, the effects of the beam depth dose contribution shall also be considered.

The beam depth dose curves for different field sizes are given in Fig. 11. From this figure, one can also observe the field size dependence of the dose drop-off. The dose drop-off was faster when the field size was smaller. For example, we could use

$$D = D_0 \exp(-\mu(x - d_{max})) \tag{2}$$

to fit the PDD curve for the depth larger than the $d_{max}$, where $x$ is the depth. It is interesting to observe that $\mu=0.006702(1/mm)$ at the field size of 6mm×6mm and $\mu=0.005935(1/mm)$ at the field size of 52mm×52mm. It was clear that the ratio was around 1.13. But the absolute values were about two orders of magnitude smaller than the DDS shown in Fig.8. The absolute value of the gradient (i.e. the first-order derivative) as a function of the depth is given in Fig. 12 for different field sizes. It was clear that the gradient of PDD was almost zero at depths larger than 1.5 cm for all field sizes. Thus the effect was negligible with a target deeper in the head. For targets close to the surface of the head, this will affect the dose distribution, such as in the case of Patient 7. Therefore, the general trend shall still hold: The DDS is larger for smaller targets or lower prescription IDLs, when the target is at an effective depth from the head surface much larger than the dmax of the beam.

## IV. DISCUSSIONS

In our work a new variable DDS was proposed to evaluate the dose drop-off in the medium-to-high dose region outside the PTV in brain SRS. The DDS was extracted from a double exponential decay fit of the relationship between the average dose in concentric shells outside the PTV and their distances from the PTV surface. Because in the double exponential decay fit the second term is much smaller than the first one by definition, one could also use the following function

$$D(r) = a_3 \exp(-b_3 r) + c \tag{3}$$

to fit the dose distribution outside the PTV. In fact, as shown in Table 2, for the double exponential decay fit $b_2$ was usually much smaller than our defined DDS $b_1$. In the single exponential function discussed here, a newly defined DDS alternative, $b_3$, shall exhibit similar dependence on prescription IDL as $b_1$ in the double exponential decay fit. It should be noted that the absolute value of the DDS depends on the plan quality. But the discovered trend that the DDS increases as the prescription IDL decreases from 90% to the lower prescription IDLs seems to be independent of the specific plans as long as the plan quality are consistent with each other for plans with different prescription IDL. One small precaution we have exercised in the work to decrease the effects of small plan quality fluctuations was excluding the first data point immediately outside the PTV from the fitting, which was proven effective. Additionally, the DDS may also depend on the target location and its surrounding structures, effects that were touched on but were not studied extensively in this paper. These effects are not expected to change the discovered DDS vs. prescription IDL relationship demonstrated in Fig. 5. The reason is that the relationship is closely related to the gradients of beam profiles of different field sizes as plotted in Fig. 10.

In understanding the normal tissue dose using the DDS, a few considerations need to be noted. First to point out is that the average dose in the shells (or rinds) was used in our analysis. Therefore, it ignored the anisotropic dependence in the relationship. However, in contrast to Dose Volume Histogram (DVH) which completely ignores the spatial information, this simple analysis still retains the spatial information along the radial direction, and can provide overall information about the dose distribution outside the PTV. But one shall keep in mind that for two different plans, even when the average dose within a rind for a plan is higher than that for another plan, point doses inside the rind can behave differently for those two plans. Secondly, although our results showed that prescription IDLs lower than 90% led to faster dose drop-off in the normal tissue near the target and hence better OAR sparing, there is one possible exception. When the PTV overlaps with an OAR, changing prescription IDL from a higher IDL to a lower one could increase the maximum dose to the OAR. This is because with a lower prescription IDL, the dose heterogeneity inside the PTV increases and therefore the portion of the OAR that is inside the PTV may likely get higher dose. Thus one needs to be careful to change the prescription IDL when the PTV overlaps with an OAR and the maximum dose of the OAR is a constraint. For example, for Patient 5 in our study, part of the brainstem is within the PTV. It was found that although the mean dose in the brainstem was the largest for the 90% IDL plan, following the trend discovered by our DDS study, the maximum dose on the other hand was the lowest compared to that in all other plans prescribed to lower IDLs.

The plans with 60%-70% prescription IDLs achieved the highest (optimal) DDS for all patients. Whilst this result indicates an advantageous normal tissue sparing in the medium-to-high dose region outside the PTV at such prescription IDLs, a few other considerations also need to be noted. Firstly and most importantly, it is well known that the probability of necrosis is higher with higher dose. As has been studied in Ref. 18, when $V_{10\,Gy}$ >10.5 cc or $V_{12\,Gy}$ >7.9 cc for normal brain tissue, hypofractionated rather than single fraction treatment should be considered to minimize the risk of brain radionecrosis. Despite the small amount of normal brain tissue contained in the SRS PTV, the higher probability of radionecrosis needs to be considered when choosing lower prescription IDLs, as the hot spots within the PTV would be considerably hotter than when a high prescription IDL is chosen instead. This potential tradeoff shall always be kept in mind when determining an appropriate prescription IDL. Secondly, as can be seen from the more spread-out lower isodose lines for the 50% IDL plan in Fig. 3, because a plan with a lower prescription IDL requires higher MUs than a higher IDL plan, the integral dose may be higher which may lead to a broader low dose spillage due to MLC and machine leakage, in addition to the higher absolute dose and higher dose heterogeneity inside the target. Although as we chose to focus in the current study, the high dose region is much more critical than the low dose region in SRS cases.  Lastly, the intention of this work is not to propose a practice pattern change to favor lower prescription IDLs in brain SRS based on our findings. Rather, the goal is to demonstrate the normal tissue dose effect of varying prescription IDLs and to explore its physical reasons. This way a physician could weigh all possible advantages and disadvantages to make an educated decision when choosing what IDL to prescribe for a specific case.

Our analysis shows general trends of dose distribution outside the PTV as the dose drops radially. Eq. (1) is a fitting function which describes the dose distribution outside the PTV for the whole imaged anatomy. This radial coordinate dependence is usually ignored in the previous measures. Therefore, the DDS can be used as a complementary measure to the previous measures. Previous measures such as maximum dose and $V_{50\%}$ are measures of special cases such as at a particular dose level, therefore could not provide the global behavior as provided in Eq. (1). Of course these measures can always be used in conjunction with our new measure to confirm the findings. For the 40 plans in our study, we also calculated $V_{50\%}$ and compared among different prescription IDLs. The results are listed in Table 3. It is clear that the same conclusion was reached based on $V_{50\%}$ as based on our proposed DDS, that the optimal sparing as indicated by the lowest $V_{50\%}$ was achieved by 70% or 60% IDL plans. Yet we need to point out again, even though the same conclusion could be drawn from both measures in this study, whilst this selection of the dose level in $V_{50\%}$ is arbitrary and local, Eq. 1 is a global fitting and the DDS extracted from it therefore provides a quantitative measure of a much broader dose range.

## V. CONCLUSIONS

A double exponential decay function was proposed to fit the dose distribution outside the PTV, from which the decay coefficient corresponding to the dose drop-off in the medium-to-high dose region was named the DDS. This new variable as an indication for normal tissue sparing was used to evaluate brain SRS plans planned with the prescription IDL from 50% to 90%. The DDS was found to increase with decreasing prescription IDLs and reach a plateau at 70% or 60%

IDL. DDS was also found to be target size dependent, smaller with larger PTV volumes. Both discovered effects can be explained by the corresponding beam profiles. .

Table Captions:

Table 1: Patient characteristics of all 8 patients.

| Patient Index | Disease | Location | PTV volume (cc) |
|---|---|---|---|
| 1 | Pituitary | Pituitary | 16.48 |
| 2 | Metastasis | Right cerebellar | 12.23 |
| 3 | Meningioma | Anterior left parafalcine | 4.37 |
| 4 | Pituitary | Pituitary | 3.64 |
| 5 | Metastasis | Right side of the pons (Brainstem) | 1.68 |
| 6 | Metastasis | Right frontal lobe | 1.05 |
| 7 | Acoustic neuroma | Left internal auditory canal | 0.56 |
| 8 | Metastasis | Right parietal lobe | 0.37 |

Table 2: The plan conformity indexes, all fitting parameters from Equation 1, and the total numbers of MUs for the SRS plans with 90%-50% prescription IDLs for an example patient (Patient 2). For this patient, the plan with 70% IDL achieved both the greatest DDS (b1), and the greatest b2, the smaller exponential decay coefficient characterising the dose drop-off in the low dose region.

| IDL | 90% | 80% | 70% | 60% | 50% |
|---|---|---|---|---|---|
| CI | 1.230 | 1.230 | 1.230 | 1.221 | 1.230 |
| a1 | 15.720 | 11.000 | 10.181 | 9.940 | 9.760 |
| b1 | 0.136 | 0.239 | **0.280** | 0.276 | 0.260 |
| a2 | 3.506 | 8.248 | 9.032 | 9.437 | 9.315 |
| b2 | 0.0371 | 0.0620 | **0.0651** | 0.0648 | 0.0634 |
| Total MU | 2465 | 2792 | 3200 | 3755 | 4531 |

Table 3: The $V_{50\%}$ (cm$^3$) calculated from the different IDL plans for all Patients.

| IDL | 90% | 80% | 70% | 60% | 50% |
|---|---|---|---|---|---|
| Patient 1 | 25.088 | 24.896 | 24.448 | 25.536 | 33.856 |
| Patient 2 | 26.752 | 20.160 | 19.392 | 20.480 | 22.400 |
| Patient 3 | 16.640 | 13.248 | 10.880 | 10.240 | 11.392 |
| Patient 4 | 13.952 | 8.256 | 7.360 | 7.232 | 7.238 |
| Patient 5 | 5.888 | 4.244 | 3.392 | 4.480 | 3.904 |
| Patient 6 | 4.416 | 2.816 | 2.624 | 2.240 | 2.560 |
| Patient 7 | 3.264 | 2.752 | 1.984 | 1.856 | 1.920 |
| Patient 8 | 2.112 | 1.280 | 1.216 | 1.152 | 1.216 |

Figure Captions:

Figure 1: The zoomed-in axial view of one example patient (Patient 4) showing multiple concentric 1mm-thick rind structures generated from the PTV surface. Although only a few rind structures are shown in the figure to avoid crowdedness, rind structures were generated layer-by-layer till they reached the closest head surface for each plan. The average dose in each rind was calculated to study the normal tissue dose.

Figure 2: One example patient plan (Patient 4 at 80% prescription IDL) showing the average dose in each rind structure as a function of the distance from the surface of the PTV. The fitted values of Eq. (1) were $a_1 = 18.43[Gy]$, $a_2 = 4.218[Gy]$, $b_1 = 0.2477[1/mm]$ and $b_2 = 0.04334[1/mm]$.

Figure 3: Dose distribution outside PTV for an example patient (Patient 2) for plans with the prescription IDL at 90%, 80% 70% (top row from left to right), 60%, and 50%. The plotted isodose lines are 20 Gy, 18 Gy, 16 Gy, 14 Gy, 12 Gy, 10 Gy, 6 Gy and 3 Gy from inside to outside of the PTV.

Figure 4: The dose distribution outside the PTV, calculated from the average doses in the rind structures, as a function of the distance from the PTV surface for the 5 plans shown in Figure 3 (Patient 2).

Figure 5: A plot of the DDS vs. the corresponding prescription IDL for all patients, obtained from all 40 plans.

Figure 6: The beam profiles for MLC- defined square fields with various sizes, generated from iPlan based on beam models for an SAD (=100 cm) setup at 5cm depth. The plotted field sizes are 6mm×6mm, 12mm×12mm, 18mm×18mm, 30mm×30mm, 42mm×42mm and 52mm×52mm.

Figure 7: The absolute gradients for the beam profiles shown in Figure 5, calculated using the first-order derivatives. The corresponding MLC defined square field sizes are 6 mm×6 mm, 12 mm×12 mm, 18 mm×18 mm, 30 mm×30 mm, 42 mm×42 mm, and 52 mm×52 mm.

Figure 8: The DDS trend with the PTV diameter, obtained from the phantom study on a simulated spherical target. The spherical target in each simulated case had a diameter of 4.90, 4.05, 3.20, 2.30, 1.20 cm, respectively. The centroid location of the varying-sized targets was kept the same and the prescription IDL was fixed at 80%.

Figure 9: The maximum DDS as a function of the PTV volume for all 8 patients. For each patient, the maximum DDS value was chosen, regardless of the prescription IDL the value was achieved with.

Figure 10: The gradients of the beam profile calculated as the first-order derivatives at 50%, 60%, 70%, 80% and 90% of the central axis dose values, for MLC defined square fields with varying field sizes ranging from 6mmx6mm to 52mmx52mm. The SAD and depth of measurement were the same as those in Figure 6 and Figure 7.

Figure 11: Percent depth dose (PDD) as a function of depth for different square field sizes. The field sizes as well as the 95 cm SSD were the same as those in Figure 6, Figure 7, and Figure 10.

Figure 12: The absolute value of the PDD gradient (the first-order derivative) as a function of depth for varying square field sizes. The gradients were calculated from the corresponding PDD curves shown in Figure 11.

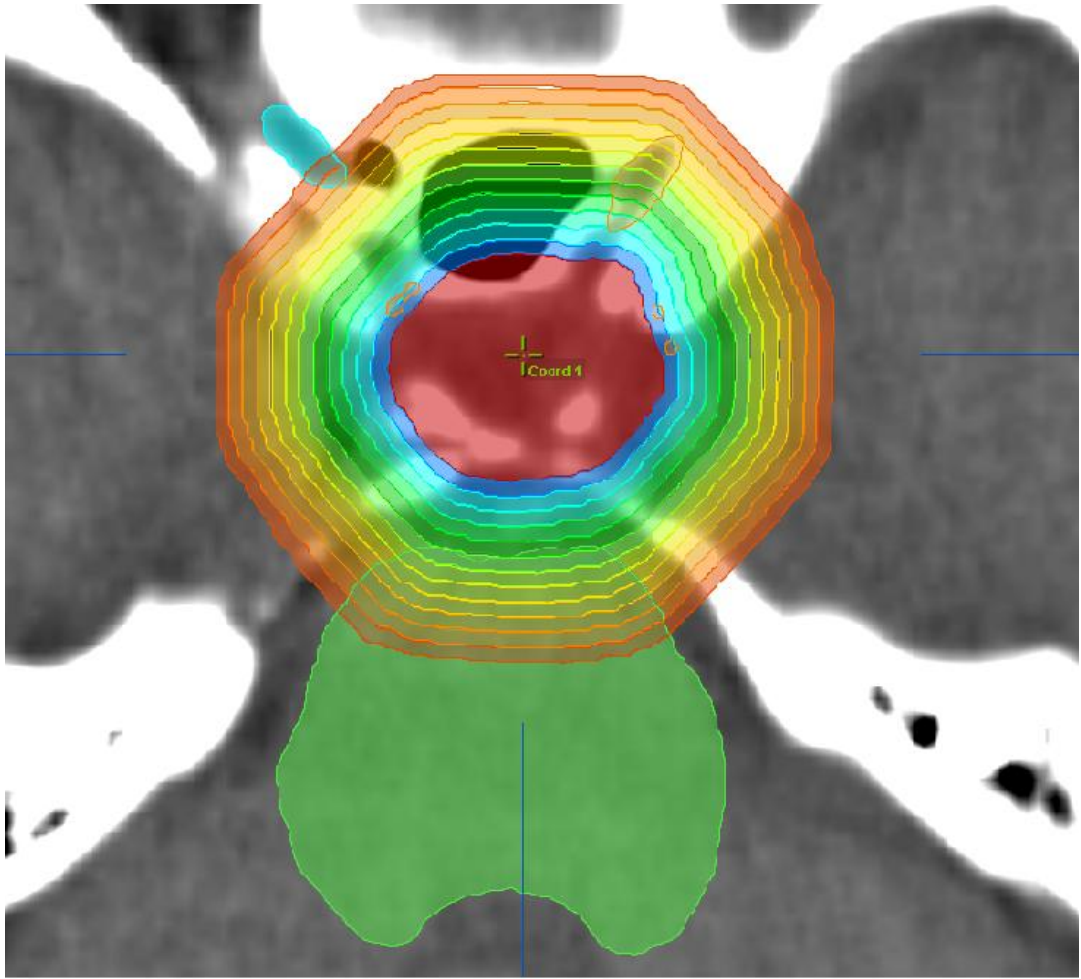

Fig.1

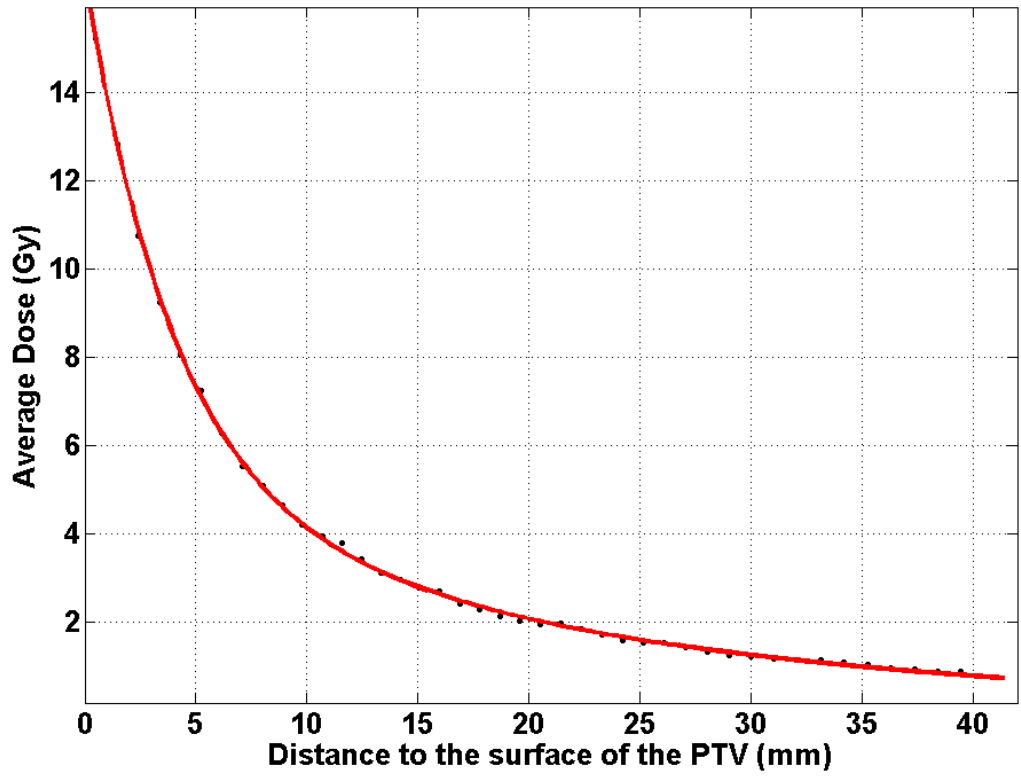

Fig. 2

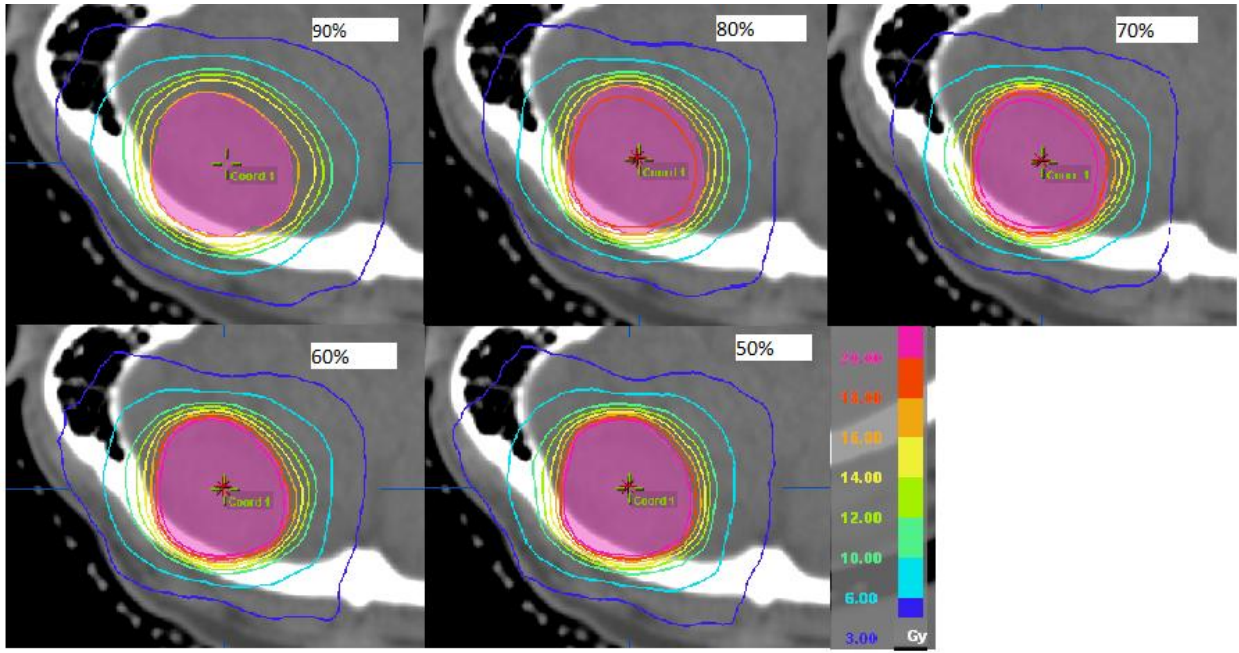

Fig.3

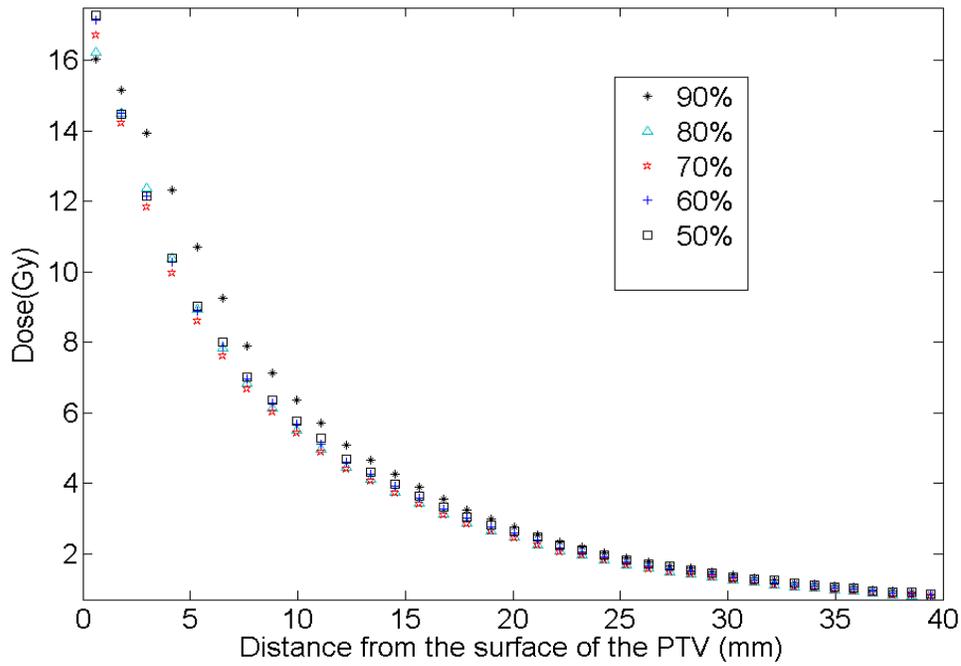

Fig. 4

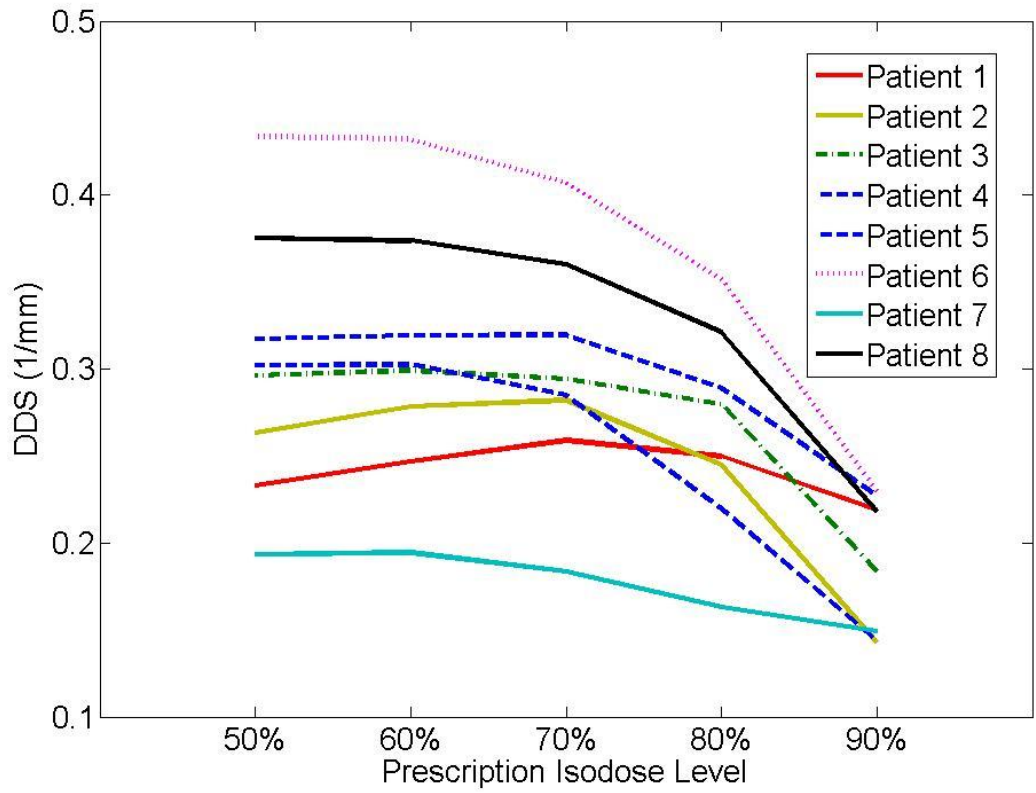

Fig.5

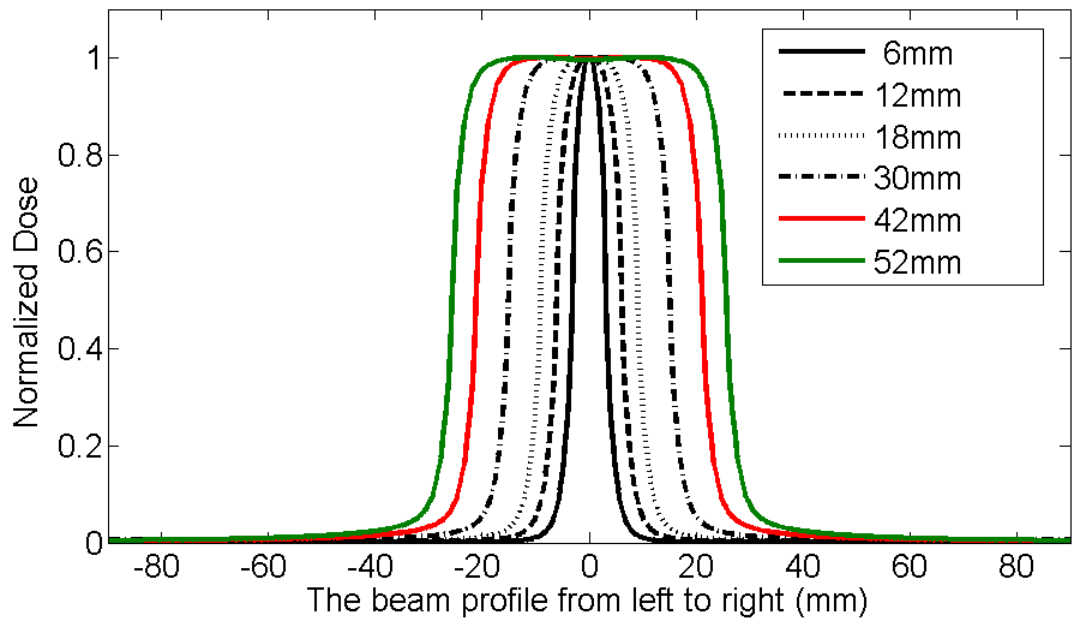

Fig. 6

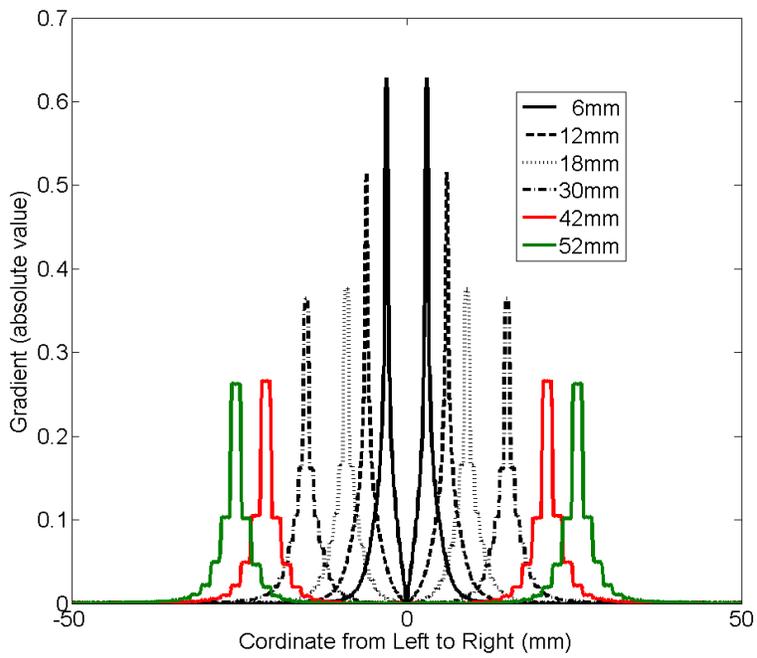

Fig.7

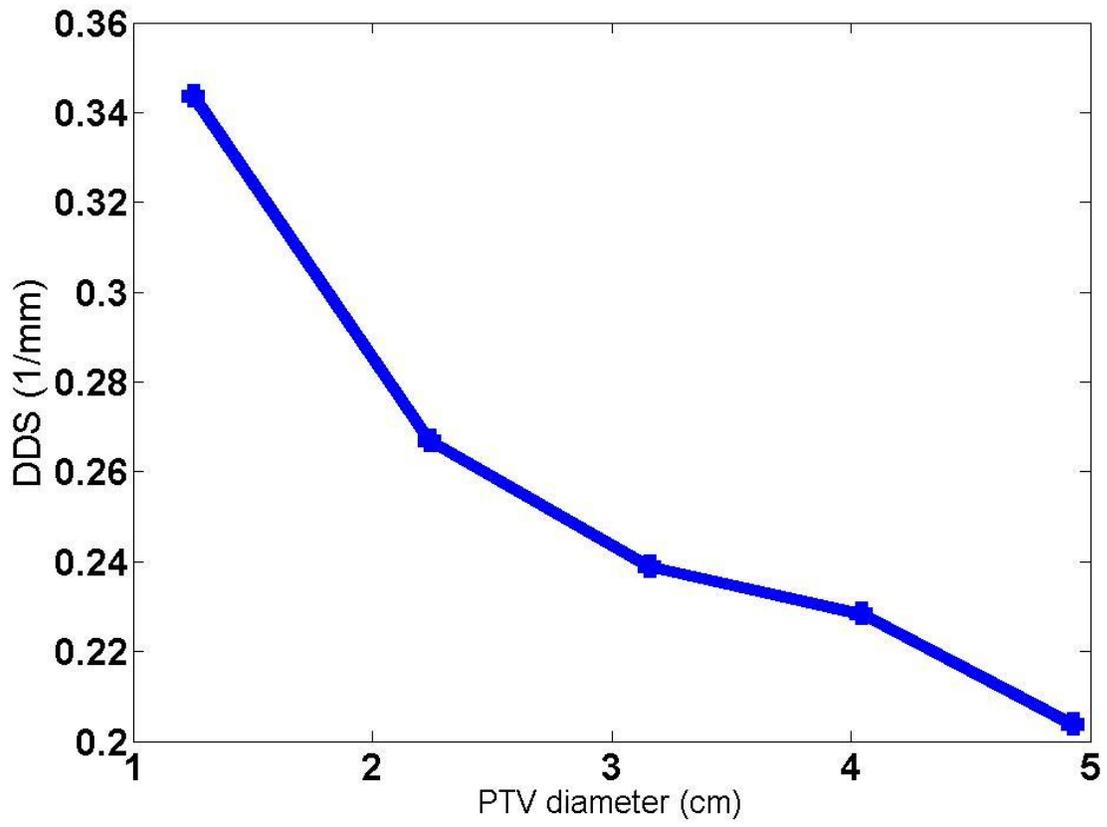

Fig.8

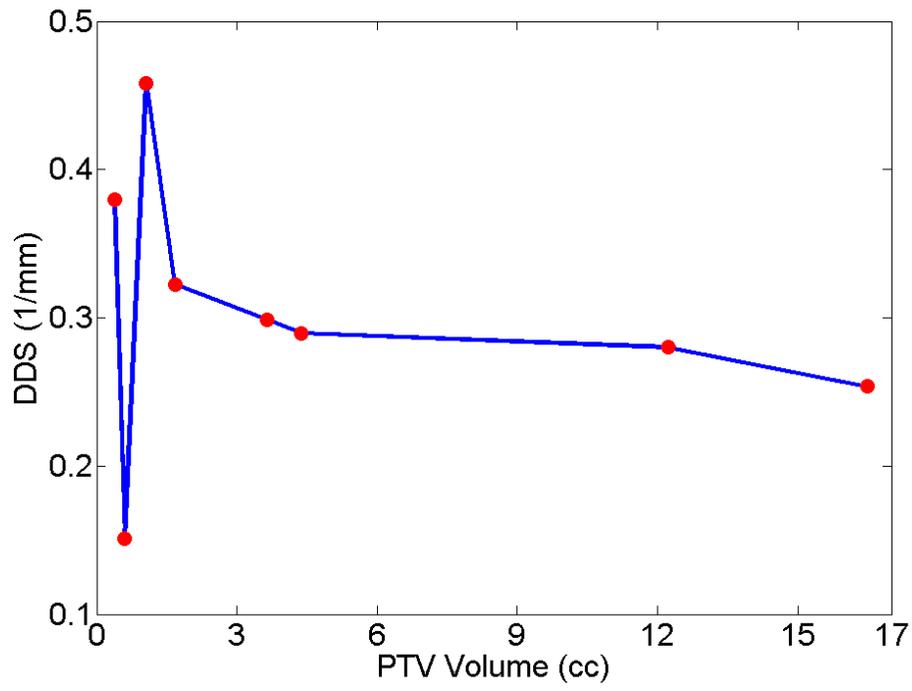

Fig.9

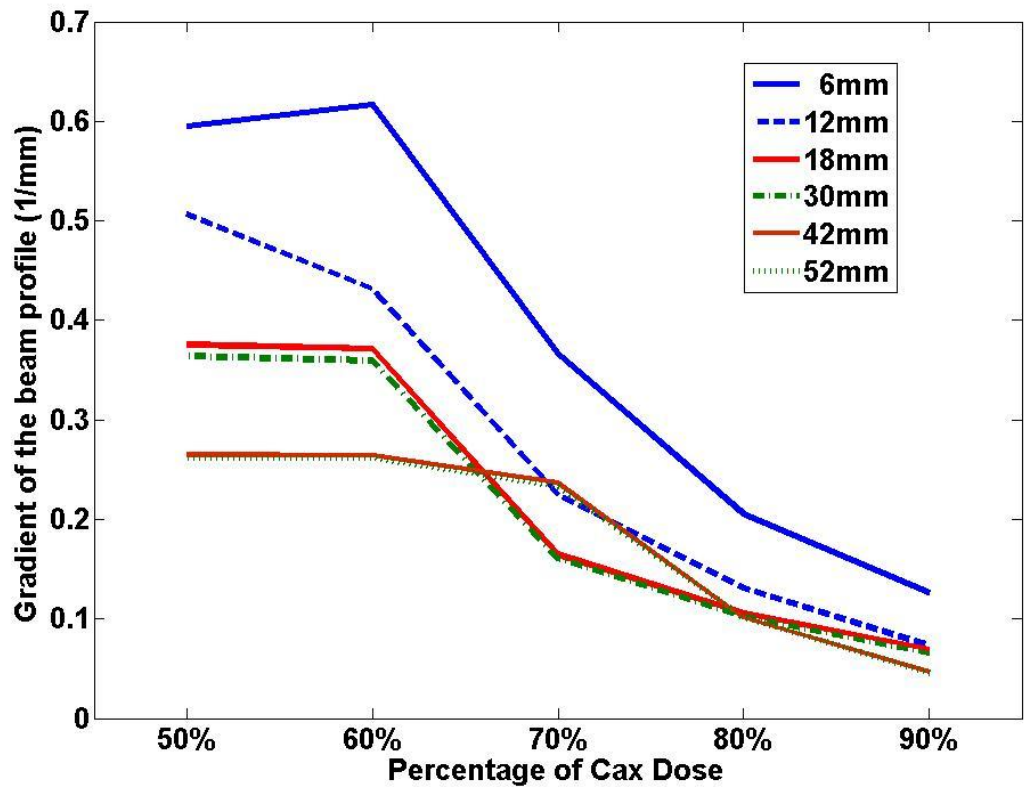

Fig.10

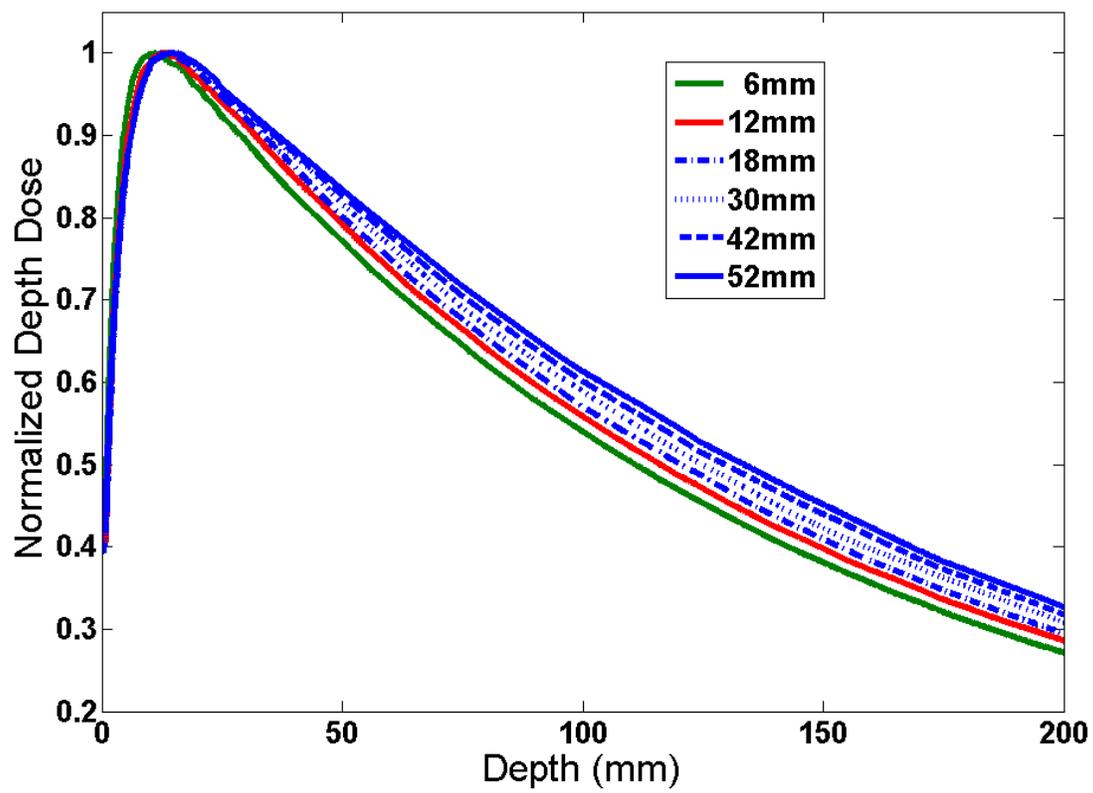

Fig. 11

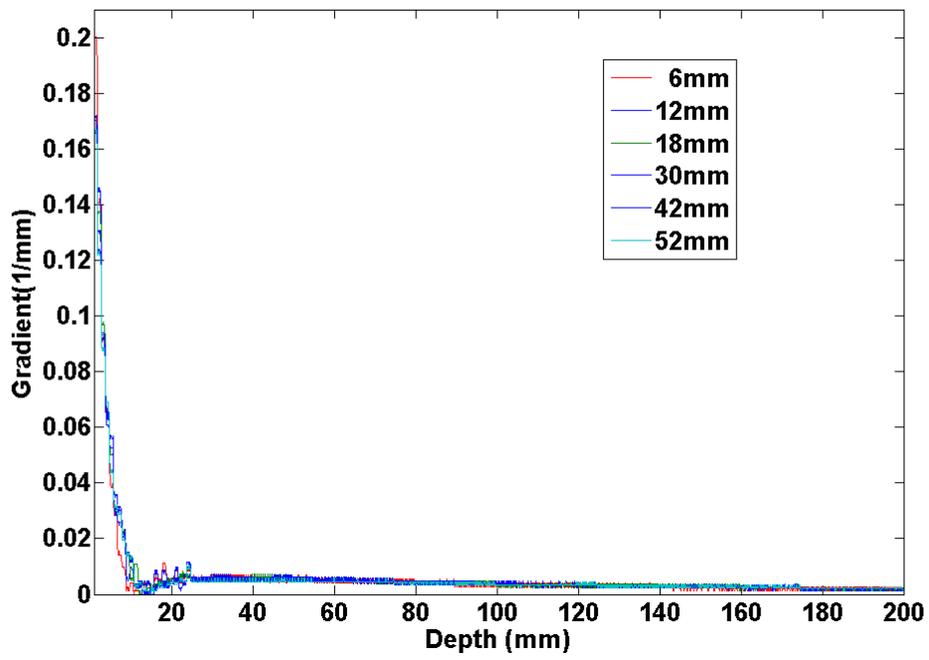

Fig. 12